\documentstyle[11pt] {article}

\title{Generalized "Quasi-classical" Ground State for an Interacting  Two Level System}
\author{
Robert Englman$^{a,b}$ and Asher Yahalom$^b$ \\
 $^a$ Department of Physics
and Applied Mathematics,\\ Soreq NRC, Yavne 81800,Israel\\ $^b$
College of Judea and Samaria, Ariel 44284, Israel\\ e-mail:
englman@vms.huji.ac.il; asya@ycariel.yosh.ac.il;}

\begin{document}
\maketitle

\newcommand{\beq} {\begin{equation}}
\newcommand{\enq} {\end{equation}}
\newcommand{\ber} {\begin {eqnarray}}
\newcommand{\enr} {\end {eqnarray}}
\newcommand{\eq} {equation}
\newcommand{\eqs} {equations }
\newcommand{\mn}  {{\mu \nu}}
\newcommand{\sn}  {{\sigma \nu}}
\newcommand{\rhm}  {{\rho \mu}}
\newcommand{\sr}  {{\sigma \rho}}
\newcommand{\bh}  {{\bar h}}
\newcommand {\er}[1] {equation (\ref{#1}) }
\newcommand{\mbf} {{ }}
\begin {abstract}
\end {abstract}
We treat a system (a molecule or a solid) in which  electrons are coupled linearly to any number and
type of harmonic oscillators and which is further subject to external forces of arbitrary symmetry.
With the treatment restricted to the lowest pair of electronic states, approximate "vibronic"
(vibration-electronic) ground state wave functions are constructed having the form of simple,
 closed expressions. The basis of the method is to regard electronic density operators as classical
  variables. It extends an earlier "guessed solution", devised for the dynamical Jahn-Teller effect
 in cubic symmetry, to situations having lower (e.g., dihedral) symmetry or without any
  symmetry at all. While the proposed solution is expected to be quite close to the exact
   one, its formal simplicity allows  straightforward calculations of several interesting quantities,
like energies and vibronic reduction (or Ham) factors. We calculate for dihedral symmetry
 two different $q$-factors ("$q_z$" and "$q_x$") and a $p$-factor. In simplified situations
 we obtain  $p=q_z +q_x -1$.

 The formalism enables quantitative estimates to be made for the dynamical narrowing of
  hyperfine lines in the observed ESR spectrum of the dihedral cyclobutane radical cation.
\section {Historical Background and Aims}

For the so called $E \otimes
e$ Jahn-Teller case (involving an electron-nuclei
system, in which a doubly degenerate electronic state is coupled
to a doubly degenerate nuclear displacement mode)
 the wave function was fully obtained as long ago as
1957 \cite{MoffitT, LonguetOPS}. The physical object of reference is
commonly a molecule of some high symmetry (say, one belonging to the cubic group, like $O$),
or a localized impurity in a solid.
The solution (or set of solutions) to this  "dynamic Jahn-Teller effect" (DJTE) are the
 vibronic states. Though this has  received, as just noted, a
full treatment early on, subsequent efforts to give simple approximate treatments or to provide
additional insight into the dynamic problem have been numerous. Descriptions of some of the
early works  are found in two books \cite {Englman, BersukerP}. Notable are the treatments in
\cite {BarentzenOO}-\cite {AvronG};  the most recent publication known to us and involving
 a variational approach to this  problem is in \cite{DunnE}.

To lead us into the present work we recall a
"guessed solution" for the ground state of the linear Jahn-Teller effect, suggested by one of
 the present authors and collaborators, which is  transparent,
intuitively simple and algebraically easily manageable. This proposal was originally worked
 out for a molecule of cubic symmetry which had a single set of interacting normal modes
 \cite{Englman62,EnglmanH, Englman}. Though not variationally obtained, the "guessed solution"
was found to have energies that are considerably closer
to the exact, computed energies of \cite{LonguetOPS} than other
approximate solutions with which it was compared. This comparison is seen
in Fig. 2 of \cite{ZhengB}. Later treatments did not test their methods
by comparison with the "guessed solution", though a critical review can be found in section
4.5.3 of \cite {BersukerP}.

The present  work is an extension of the earlier approach to a substantially broader and harder
problem, namely
to a pair of electronic states in unrestricted symmetry and subject to interaction with an arbitrary number
 of nuclear displacement modes, but only in a linear manner.
   The subject of two-state interacting with bosons (which may either be phonons or photons)
  has had a very extensive literature. The spin-boson Hamiltonian that forms the starting point of
  \cite {Leggett} is a
  special case of the Hamiltonian introduced in this paper. Likewise, several
  books contain accounts of the related Jaynes-Cummings method \cite {Optics,Optics2}. The
    present work also belongs to this field, but is restricted to a pair of ground
    level states. Even with this restriction, the closed solution that we present here can find
    its uses in treating the energy dissipation of a spin system \cite {Leggett}.

     The handling of  external perturbation {\it after} taking care of the electron-nuclear
     interaction is a potential tool to tackle Berry phases in open systems \cite {CarolloFFV,
     WhitneyG}.
    We would also recall a recent work on the Jahn-Teller effect in lower
     than cubic symmetry, which is less general than the present one, but has permitted us to
     check some of our results numerically \cite {MajernikovaS}. The reduced symmetry case (named
     "the elliptic form" to differentiate it from the circular energy trough in $E \otimes
e$) was studied previously in \cite{BaerME}. We calculate (for the first time, to our knowledge)
the experimentally important reduction factors for the low symmetry case (section 4.1), having
pointed out (at the end of section 3.4) that, when the electron-nuclear coupling is strong,
one meets broken symmetry instabilities.

The formalism, initially formulated in very general terms, is gradually shifted to more specific
situations, such as systems of cubic and of lower (e.g., dihedral) symmetries, and to systems
 with two (rather than an arbitrary number of) vibrational modes and, ultimately, to a specific molecular
 system. In this last, the formalism and the numerical results for the reduction factors lead
to  quantitative conclusions for
the dynamical narrowing of  hyperfine lines in the observed ESR spectrum of the dihedral
 cyclobutane radical cation. This is the subject of section 5.2.

\section {A General Hamiltonian}
We now write down a  Hamiltonian for a pair of (diabatic, ~or nuclear
coordinate-independent) electronic states, denoted by the symbols $ \zeta_{\theta}$ and $\zeta_
{\epsilon}$). The states  are understood to be functions
 of any  number of electronic coordinates, e.g, all the electrons in an atom or a molecule,
 but this functional dependence is absent in the formalism as long as the behavior of
  the doublet states alone is under consideration. A two-state situation can come about for an atom
 that is placed in a strongly coupled environment, such that this separates the doublet
 from the rest of the
  electronic manifold, or for a molecule in which the internal, intramolecular forces achieve
  the same effect. The two states need not be degenerate but, in order that it should be legitimate
  to consider them separately, they must be, in some sense, isolated
   from the rest of the electronic states (e.g, either by symmetry consideration or by a large energy
  gap).  If so, then the "two level-the rest" matrix elements of all interactions can
    be neglected to some approximation. This is the physical setting for the formalism that follows.
    It leads naturally to the  representation of the
    electronic states as the column vectors \beq ( \zeta_{\theta},  \zeta_{\epsilon})~=~(\left( \begin{array}{cc}
  1 \\
  0
  \end{array} \right),\left( \begin{array}{cc}
  0\\
  1
  \end{array} \right))\label {zeta}\enq

  The electronic states are coupled to any number of nuclear
  displacements coordinates $q_n ~(n=1,...N)$. We assume that these are organized into a set of
   normal modes brought to a standard form (i.e., having the same
  effective mass) and  restrict the coupling to be
   of no higher order than linear in the displacement coordinates. Thus, one has for the displacement coordinates
   the following harmonic oscillator Hamiltonian:
  \beq H_{nuc}=\sum_{n=1}^N \frac{\hbar\omega_n}{2}(-\frac{\partial^2}{\partial q^2_n} + q^2_n)
  \label{HO}\enq
  where $\hbar\omega_n$ are the quanta of vibrational energies. In the two-state representation
   the nuclear Hamiltonian is written as a scalar or, equivalently,
  as $H_{nuc}$ times the $2$x$2$ unit matrix $I$.

   The remainder of the Hermitian
  representation matrices for the two level system are the familiar Pauli-matrices
  \beq \vec{\sigma}=(\sigma_x,\sigma_y,\sigma_z)=(\left( \begin{array}{cc}
  0 & 1\\
  1 & 0
  \end{array} \right),\left( \begin{array}{cc}
  0 & -i\\
  i & 0
  \end{array} \right),\left( \begin{array}{cc}
  1&  0\\
  0 & -1
  \end{array} \right))\label{sigmas}\enq
  In terms of these we can write out a general linear form of interaction between the
  electronic motion and the (real) nuclear coordinates (in the absence of any molecular
   symmetry), as
  \beq
  H_{el-nuc}= -\sum_{n=1}^N \frac{\hbar\omega_n}{2}(a_nq_{n}
  \sigma_z +b_nq_{n}\sigma_{x})
\label{Hen}\enq
in which (the dimensionless) $a_n$  and $b_n$  express the strength of interaction
  between the electrons  and the nuclear motion in the $n$- mode. (A detailed
  discussion of the linear many mode interaction in a symmetrical setting is found in section 3.5.3
   of \cite{BersukerP}. In \er {Hen} frequency changes between the two states are ignored to
   be consistent with a purely linear coupling.)

    $\sigma_y$ is absent in the above interaction
   Hamiltonian, as also in several previous works \cite{MoffitT}-\cite{ThielK}. When the states
   of the two level system are orbital states then, for all molecular point
   groups considered in this work, the symmetric product of the state-representations
   does not contain the representation of the $\sigma_y$-matrix. When the two states are
   a Kramers-doublet, the situation becomes more complex, since in a low order perturbation
     the coefficients $a_n$  and $b_n$  vanish, unless some further effects
     (like crystal field, spin-orbit interaction, external magnetic fields ${\vec H}$,
     spin-spin coupling) are included in the perturbational calculation of these coefficients.
Working out the spin-lattice coupling for a Kramers' doublet (on a six coordinated $Cu^{2+}$)
Stoneham gave symmetry arguments (in the last equation of \cite {Stoneham}) to show
that, for both $a_n$  and $b_n$ to be non-zero, both  $H_x$ and $H_z$ need to be non-vanishing
(while $H_y=0$). This corresponds to the form shown in the above equation (where the coefficients
would be magnetic field dependent). If, on the other hand, $H_y$ is
also non-vanishing, then for a Kramers' doublet a term with $\sigma_y$ will also be present.
Since this term is absent in orbitally two-state system, we do not complicate the formalism
by adding the $\sigma_y$ term.

 One also has to consider the electron being acted upon by
 external fields. (The interaction of the external field on the nucleus is supposed to be
 contained in the potential of the nuclear coordinates.) In the preceding,
  vector representation of the two states, any interaction Hamiltonian (that expresses
  the coupling between the electron and any external field) must have the form
 \beq
H_{el-f} =\frac{\hbar}{2}{\vec \Omega }\cdot\vec{\sigma}\label{Hef}\enq  with the
 representative of the fields ${\vec\Omega}  =(\Omega_x,\Omega_y,\Omega_z)$ inside the
 two level system being constant (independent of the values of the electronic or of the nuclear
variables), this being the most general form of expression for the system. In section 4,
which discusses the effect of external forces, we give examples for the interaction.

Any difference between the two-state energies can be considered to be part of
$\Omega_z$ so that, until we come to the subject of the external fields in section 4,
the states can be considered as a pair of {\it degenerate} doublets. However, the rest
 of $\Omega_z$ (as well as $\Omega_x$ and $\Omega_y$) comes from externally applied sources.

The total Hamiltonian $H_{tot}$ is the sum of the previous Hamiltonians
\beq H_{tot}=(E_0+H_{nuc})I+H_{el-nuc}+H_{el-f} \label{TotalH}\enq
to which has been added a scalar term  with $E_0$ representing the mean energy of the
non-interacting states.
(The spin-boson Hamiltonian which forms the basis of \cite {Leggett} is obtained from \er{Hen}
 and \er{Hef} upon putting
 $a_n\ne 0$, $b_n=0$,  $\Omega_x \ne 0\ne \Omega_z$,$\Omega_y=0$.)

The treatment of $H_{el-f}$ will be postponed to later. In its absence,
 we have a  pure "vibronic (=vibrational-electronic)" situation, which we now treat.

 \section {Vibronic doublet}
 The Hamiltonian
\beq H_v=H_{nuc}+H_{el-nuc}\label{Hv}\enq
involving (partially) the electronic and nuclear degrees of freedom will be the subject
of our investigation in this section. We first show
 that solutions of the partial Hamiltonian form a degenerate {\it doublet} (with the understanding
  that the energy difference between the two states is shifted to the external field part).
 \subsection {State degeneracy}
 The following non-identity transformation leaves the above Hamiltonian invariant:
  \beq {\cal T} = [{\cal A}_e][{\cal P}_q]\label{T}\enq where ${\cal A}_e$ stands for the
   following simultaneous
  changes in the electronic states $\left( \begin{array}{cc}
  1 \\
  0
  \end{array} \right)$$\to$ $\left( \begin{array}{cc}
  0\\
  -i
  \end{array} \right)$ and $\left( \begin{array}{cc}
  0 \\
  1
  \end{array} \right)$ $\to $ $\left( \begin{array}{cc}
  i\\
  0
  \end{array} \right)$ and ${\cal P}_q$ is the parity operator for all mode coordinates,
  namely
  \beq P_q q_n =-q_n ~~~n=1,...N \label{parity}\enq The transformation ${\cal A}_e$ can be
  achieved by the unitary matrix $\sigma_y$. But, clearly,
  \beq {\cal T}^2= 1 \label{tsquare}\enq so that, simultaneously with any eigenvalue of
  the Hamiltonian $H_v$, ~${\cal T}$ has an eigenvalue $+1$ corresponding to a state
   unchanged under the ${\cal T}$-transformation and another eigenvalue $-1$, that
   corresponds to (another, different) state that changes sign under this transformation.
   The conclusion is that the vibronic Hamiltonian has doubly
    degenerate  eigenstates. This degeneracy is lifted, when
    field interaction term $H_{el-f}$ is inserted, as we shall see.
\subsection {The quasi-classical ground state}
Based mainly on the numerical agreement of the energies under $O$ symmetry, noted in the
 opening section, we extend  here the method of \cite {Englman}, \cite {EnglmanH} to
 the general case under study,  that is, we propose the following form for the ground state wave-function
\beq {\hat\Psi}= {\cal N'} \exp\{-\frac{1}{2}\sum_{n=1}^{N}(q_n I-\frac{a_n}{2}\sigma_z -
\frac{b_n}{2}\sigma_x)^2\}\label
{Psihat}\enq
where $\cal N'$ is a normalizing factor. The wave function {\it generator} ${\hat \Psi}$ is
a matrix (or operator). It possesses the
full ($A_1$) symmetry of the Hamiltonian. Therefore operating with ${\hat\Psi}$ on an electronic component with some symmetry,
  will generate a state with the symmetry of the component. In order to get the probability amplitudes in
  the two electronic states of \er {zeta}, one has to left-operate with
   ${\hat \Psi}$ on the basic vectors or on any linear combination of them, as will be shortly
   described. The prescription (and the underlying rationale) for the proposed construction
  is to regard the Pauli matrices (which are the electronic density operators) as
   c-numbers. Having done this, we write down the ground state
   wave function in the form of a set of displaced independent vibrational coordinates. The
   initial handling of (quantum mechanical) matrices in the manner of c-numbers has suggested
    naming the method "quasi-classical".
      However, the modes are now no longer independent: thus the moments (e.g., the expectation
   value or the spread) of any mode depends on the coupling constants of the other modes.

     The mathematical meaning of the exponential form in
   \er{Psihat} is that one has to expand the exponential in a power-series of the exponent.
   Remarkable in the posited ${\hat \Psi}$ is that the individual frequencies do not appear in
   it (just as they do not in the wave function of a set of uncoupled oscillators, when expressed in
   a standard form). Of course, the energy expectation value depends on the frequencies, since
    the Hamiltonian does and so do, implicitly,  the dimensionless coupling constants $a_n$ and $b_n$.

   The success of the method hinges on the fact that it is possible to sum the power
    series exactly, in spite of the non-commuting terms in the exponent. This is made possible by
     the property of the two dimensional spin (Pauli) matrices, that
    \beq \sigma_i\sigma_j +\sigma_j\sigma_i ~=~2\delta_{ij}\label{sigsig}\enq where $\delta_{ij}$
    is the Kr\"onecker delta. Using this property, one readily obtains the simplified expression
 \beq{\hat\Psi} ={\cal N} [I\cosh \frac{e({\vec q})}{2}+ \frac{\sum_{n=1}^N (\sigma_z a_n q_n
    +\sigma_x b_n q_n) }{e({\vec q})} \sinh \frac{e({\vec q})}{2})]
   \label{Psihat2}\enq with ${\cal N}$ being another normalizing factor.
   In the argument of the hyperbolic functions one has
   \beq e({\vec q})=  \sqrt{[(\sum_{n=1}^Nq_n a_n)^2 +(\sum_{n=1}^{N}q_n b_n)^2]}\label{e1}\enq

   It must stressed again that the resulting quasi-classical wave-function is
   only an approximation, whose accuracy depends on how well the Pauli-matrices can be
   approximated by c-numbers. This will be the case when e.g., one of the potential wells is
    deep, since then the energy of the state will be dominated by the electron occupancy near
    the minimum.  On the
    other hand, when the frequencies of the different oscillators differ markedly, the
    quasi-classical approximation could be in error, since then, e.g., the positions of the
    saddle points in the potential may not coincide with the maxima in the overlap of
     wave-functions coming from different potential wells. A scale transformation applied
     to each well, in the form proposed in  \cite {DunnE, LiuBDP, LiuDBP} for cases of
     higher degeneracies  than two and  modes of various dimensionalities,
      could lead to improvements in the wave-function, but requires a formalism that
      is more complex than the one advocated here.
     \subsection {Some elementary symmetry considerations}
As already emphasized, there need not be any relation (symmetry-based or otherwise) between the
two states and among the nuclear coordinates for the foregoing formalism to hold.
However, if the system has some symmetry properties (or we choose to relate it to a symmetric
framework) things become at the same time clearer, more systematic and more familiar.

We therefore formulate the foregoing in a symmetry-setting and employ implicitly the theory
of point molecular or point-groups.
\subsubsection {Cubic symmetry}
In a system which nominally belongs to a
 cubic symmetry group (like $O$), the two electronic states could belong to a doubly degenerate
 E representation, whose two components are designated in \cite {Griffith} as ($\theta,
 \epsilon$). This designation was already used by us above in \er {zeta}. We have chosen these
 symbols in preference
 to others, such as those in \cite{KosterDWS}, because we shall later use the so-called W-coefficients
  and these were tabulated for the symmetry groups of interest in  \cite{Griffith} using the
  present symbolism.
 In the expression, \er{Hen}, for the coupling to the nuclear motion coordinates $q_n$, the coefficients
  $a_n$ are non-zero
 for modes belonging to the $\theta$-representatives of a two-fold $e$-mode and $b_{n}$ are
  non-zero  for modes belonging
  to the $\epsilon$-component of an $e$-mode, with the two coefficients  being numerically
  pairwise equal. This is, of course, the multi-mode $E\otimes(e_1+e_2+...+e_{N/2})$
Jahn-Teller situation, described in detail in \cite{Englman},\cite{BersukerP}.

 ${\hat\Psi}$ in \er{Psihat}  generates vibronic wave function in the following way. When
  it is let to operate on $\left( \begin{array}{cc}
  1 \\
  0
  \end{array} \right)$, one obtains the $\theta$- component of the ground state vibronic
   doublet (in the present quasi-classical approximation); if it is let to operate on $\left( \begin{array}{cc}
  0\\
  1
  \end{array} \right)$, one obtains the $\epsilon$-component of the same. For future use
  we write these {\it vibronic} wave-function components in a curly ket form, as
  \beq{\hat\Psi}\left( \begin{array}{cc}
  1 \\
  0
  \end{array} \right)  =  |\theta\},~~~{\hat\Psi}\left( \begin{array}{cc}
  0 \\
  1
  \end{array} \right)  =  |\epsilon\}\label{vibronics}\enq

  If, instead, one operates with ${\hat\Psi} $ on the two orthogonal
   linear combinations  $\frac{1}{\sqrt(2)}\left( \begin{array}{cc}
  1 \\
  \mp 1
  \end{array} \right)$, one reaches a pair of other vibronic states, preferentially
  localized
  in a different part of the coordinate space than the ones in \er{vibronics}. (These vibronic states have properties
   similar to polaronic states, which term is in use for a single electronic state.)

    If, alternatively, one operates on the following  complex combinations of the electronic
    states,
  $\frac{1}{\sqrt(2)}\left( \begin{array}{cc}
  1 \\
  \mp i
  \end{array} \right)$, one obtains vibronic states that have values of $+1$ and $-1$
   of a "composite" vibrational angular momenta, this being defined in terms of the
   composite vibrational angle variable, given by
   \beq\Phi=\arctan \frac{\sum_{n'=1}^{N} b_{n'} q_{n'}}{\sum_{n=1}^{N} a_n q_n}
   \label{Phi}\enq
    Clearly, although the vibrational modes were originally independent, the resulting
    angular  variable is not the sum of the mode angular variables, like
   \beq \Phi=\sum_{m=1}^{\frac{N}{2}}\sum_{m'=\frac{N}{2}+1}^{N} \arctan \frac{q_{m'}}{q_m}~~~~(not~true)
  \label{Phin}\enq  This is, of course, due to the coupling of the modes to the
    electronic degree of freedom.

    A pair of states in any of the combinations are energy-degenerate and
    mutually orthogonal.
\subsubsection {Dihedral symmetry}
    In a lower symmetry situation, like $D_{2d}$, the doublet state could belong to the doubly
    degenerate $E$-representation, with components designated by ($\theta,\epsilon$),
    as before. The $a_n$ coefficients are non-zero for modes possessing $b_2$ symmetry and
the $b_{n}$ coefficients belong to $b_1$ symmetry types. However, this time (unlike for
O-symmetry) there are no symmetry-based relationships between $a$'s and $b$'s.
The operator ${\hat\Psi}$ still possesses
full ($A_1$) symmetry. This can best be seen in the form given in \er{Psihat2}. Here the
first term is clearly invariant under all group operations (since the squares of these
are the identity
 operator) and the second term is also invariant because it has the symmetry of the
  Hamiltonian. Therefore operating with ${\hat\Psi}$ on an electronic component with some symmetry,
  will again give a state with the symmetry of the component.
 \subsection{Energies and other expectation values in dihedral symmetry.}
 The advantage of the quasi-classical form is that the expectation values of c-numbers or
 of operators can be calculated using the explicit form of the states just given. (A recent
calculation of the expectation value of an angular momentum {\it operator} is in \cite {EnglmanYPRA}.)
 As already
  noted in the opening section, these expectation values have proven to be very accurate for the single mode
  case in O symmetry, where comparison was made with the exact, computed results that were
  available.
 One can  carry out  similar calculations for several modes
  in any dihedral group, like one of $D_4$ symmetry (as well as in other dihedral groups,
  like $D_{4h}$, $D_{2d}$, etc). Actual results will not be given here for several modes,
   since these will
  depend on details of the system. (Some relevant molecular systems will be considered
   in section 5.)

   A sporadic comparison has been made with results of a recent paper \cite {MajernikovaS}.
    This paper showed graphs of
    eigenenergies of the (degenerate) vibronic ground computed exactly (by a numerical method),
    as well as with several approximation schemes. We compare the energy expectation values
    computed within our quasi-classical
    approach with theirs, at values of parameters for which the discrepancies between the
    exact and approximate values appear to be largest. This is the region where neither
     perturbation theory (weak coupling), nor asymptotic formula (very strong coupling)
     holds. Relating to  figures 3(a) and (b) in \cite {MajernikovaS} and to parameter values $a=2,
      b=1.5 $ (equal
     to $\mu=0.5, \chi= 0.75$ in the symbols of \cite {MajernikovaS}), their exact eigenenergy
    is $E=-0.16725$. We compare this to $-0.15075$ obtained by our
    method, which is higher (as it should be for a non-exact expectation value)
       by $0.0165$. This discrepancy         is
       worse than the {\it best} approximation in \cite {MajernikovaS}, rather better than
       the next best (obtained variationally) and considerably better than three others. Testing additionally
       our quasi-classical approximation against the exact results exhibited in the figures 4 (a) and (b)
     of the above reference, for $a=2\sqrt 2, b= 1.5\sqrt 2$ ($\mu=1.0, \chi= 0.75$), the exact
      value is
     $E=-1.17295$, with which we can compare our value of $-1.13370$, or a discrepancy of $0.03925$.
     This is somewhat worse than the first and second best approximations of \cite {MajernikovaS}.

     In summary, it seems  that the intuitive, quasi-classical method can answer most
      practical needs for energy values. One would expect at least semi-quantitative guidance
      for other numerical quantities,when derived from the quasi-classical model. The next section
      contains some such quantities.

  It is also of interest to consider the limiting case of very strong electron-vibrational
  coupling. This comes about when at least some of the coupling strengths $a_n$ and $b_{n}$
  in \er{Hen} are numerically much larger than unity. (The opposite extreme of zero coupling
   has for its "vibronic" ground states the product of gaussian states multiplying some linear
   combination of the electronic basic states.)

   In the strong coupling limit the vibronic ground state is  stabilized by the
   coupling and again takes the form
   of a product of vibrational and electronic factors. The form of the latter depends critically
   on the ratio of two stabilization energies. (This result goes back to \"Opik and Pryce's classic
   paper \cite {OpikP}. Modifications due to higher order coupling are treated in \cite
   {Bacci}.) The stabilization
    energies (analogous to the static Jahn-Teller energy \cite {Englman,    BersukerP}) are
    \beq \Delta E_{b_2}= \frac{1}{8}\sum _{n=1}^N \hbar \omega_n(a_n)^2\label {stab1}\enq
    \beq \Delta E_{b_1}= \frac{1}{8}\sum _{n=1}^N \hbar \omega_n(b_n)^2\label {stab2}\enq
    where the subscripts $b_2$ and $b_1$ signify the types of the distortion mode
    in $D_{2d}$ symmetry. The ratio of the two
   \beq R=\frac{\Delta E_{b_2}}{\Delta E_{b_1}}\label{R}\enq enters now so that for $R>1$
   the stabilized electronic states are the $(\theta, \epsilon)$ components, shown in \er{zeta},
    and the stabilization energy  is $\Delta E_{b_2}$, while for $R<1$, the stabilized
     electronic states are $\frac{1}{\sqrt(2)}\left( \begin{array}{cc} 1 \\
  \mp 1
  \end{array} \right)$ and the stabilization energy is $ \Delta E_{b_1}$. ($R=1$
  is a coincidental case in dihedral symmetry. It is the normal case under cubic symmetry and
  results in a continuum of stable configurations, rather than a single point in the
  multi-dimensional configuration space for each state.)

  The transition between  $R<1$ and $R>1$ represents a change from one broken symmetry
  type to another. The subject of broken symmetries is, of course, an important issue for
   macroscopic systems and also for the observed inhomogeneous state of the universe. In these
    cases each stabilization energy is large on a characteristic
    quantum scale (in a macroscopic or in an astronomical manner).
   Yet, it is a tiny difference between the stabilization energies that tips the balance between
   different types of broken symmetry. This means that if, in the neighborhood of a situation
   where the stabilization energies are the same, the coupling constants are made varied, then changes
   in the macroscopic symmetry and energy can come about by microscopic causes. (This is
    exemplified in our treatment of the cyclobutane radical cation in section 5.2.2.) The  foregoing
   treatment is of course very approximate and does not take into account higher order
    couplings or thermal fluctuations; still, as a model it is instructive and can prove to be
    helpful for extensions to more realistic cases.

\section {External Fields}

The form of the interaction, shown in \er{Hef}, between the electronic part and an external
 field (represented by $\vec {\Omega}$)
is the consequence of the Hermitian nature of the Hamiltonian, the three matrices
 $(\sigma_x,\sigma_y,\sigma_z)$ being the only $2$x$2$ matrices having this property
  (apart from the unit matrix, which only shifts both states by an equal, constant amount).
\subsection {Examples for ${\vec\Omega}$}
  The simplest example is the Zeeman effect on an electronic spin, for which $\vec {\Omega}=\beta
  {\vec H}$ (with $\beta$ being the Bohr magneton). The effect of a magnetic field $H_z$ acting
  within the two states ($t_{2g,\xi}, t_{2g,\eta}$), which are split off from a $3d$-state
   manifold by crystal fields of cubic and tetragonal symmetries, would be represented by
   $\Omega_y=$constant$~H_z$, where the constant includes $\beta$ and a radial integral (\cite
    {Stoneham}, Table 1). A uniform stress of
 the type  $\tau_{zz}$ acting upon a doubly degenerate set under $C_{2v}$, with $z$ the fourfold axis.
 will be expressed by ${\Omega_z}$. An applied variable electric field given
 by a potential $V({\vec r}) =  xyf(|{\vec r}|)$, when acting within an electronic pair
  having the real forms $|1>= xg(|{\vec r}|)$, $|2>= yh(|{\vec r}|)$ will yield \beq {\Omega_x}=
 <1|V({\vec r})|2>\label{V1}\enq with all other $\Omega$-components being zero.

 In a general way, for a perturbational Hamiltonian ~$\Delta H$ being an arbitrary function of
 the coordinates, the $\Omega$- magnitudes and real and imaginary parts of the matrix
  elements inside the $1,2$ manifold are connected by
 \beq \Omega_x=Re (\Delta H)_{1,2},~~\Omega_y=-Im (\Delta H)_{1,2},~~\Omega_z
 =\frac{1}{2}((\Delta H)_{1,1}-(\Delta H)_{2,2})\label{omegas}\enq

Since the interaction term $H_{el-f}$ leads naturally to consideration of the vibronic, or Ham-, reduction
factors \cite {Ham1,Ham2}, we introduce these now.

\subsection {Reduction factors}
One starts with some coupling affecting the electrons. Within the two state manifold, this
 can be expressed in terms of matrix elements between these (adiabatic)
 states. The question is how do these  matrix elements change, when the two states are  no longer
  purely  electronic, but rather  vibronic (coupled electronic-vibrational) states? The
  answer, given in a context similar to the present one and originally due to Ham \cite
  {Ham1,Ham2}, is a reduction in the strength of the original coupling by factors originally
  denoted by $q$ and $p$ (that are $1$ in the absence of vibrational coupling and less than $1$
  in their presence). These reduction factors
 have further been described in \cite {Englman} and, at considerable length, in \cite {BersukerP}
 where references to several literature sources can be found. Section 4.7 in the book \cite
  {BersukerP} contains an analysis of the
 reduction factors in terms of W-coefficients, tables for which (in both cubic and dihedral
 groups) can be found in \cite {Griffith}.

In the lower symmetry situation  under concern here (e.g., in dihedral symmetry), when the
coupling strengths in the diagonal position differ from
those in the off-diagonal positions, there are three reduction factors. These are here named
$q_x$, $p$ and $q_z$.
The formal definitions are in terms of the scalar product of the vibronic states
introduced in  \er{vibronics}, in the form
\beq q_x  =  \{\theta|\sigma_x|\epsilon\}  =  \{\epsilon|\sigma_x|\theta\}\label
{qx}\enq \beq q_z  =  \{\theta|\sigma_z|\theta\}  =   -\{\epsilon|\sigma_z|\epsilon
\}\label{qz}\enq \beq p  =  i \{\theta|\sigma_y|\epsilon\}  =  -i \{\epsilon|\sigma_y|
\theta\}\label{p}\enq
(On notation: The subscripts of the $q$'s agree with those of the $\sigma$-matrices; the letter
$p$ has been retained in preference to a possible $iq_y$ for historic reasons. In terms of the more recent
symbols $K(a)$ used in, e.g., section 4.7.1 of \cite {BersukerP}, where $a$ is a
representation of the group, one can identify:
\beq q_x= K(B_2)~,~q_z= K(B_1)~,~p= K(A_2)~, ~1=K(A_1)\label{pK}\enq
The symbol $K(a)$ will be used later in obtaining formal expressions for the reduction
factors.
 In a higher symmetry situation (e.g. $O$) $q_x=q_z=q$, as shown on p.43 of \cite{Englman},
  where the ket in Eq. 3.38  should be corrected to be an $\epsilon$-type.)
 The results of our computations are shown for the simplified situation that there is a single
  $b_2$ and a single $b_1$ mode coupled to the electrons. When the coupling strength are
  of the same strength ($a=\pm b$), our results  are identical to those quoted in the literature
  for $O$-symmetry. In particular,
  \beq q_x=q_z=q, ~~ p=2q-1 ~~(in~O~ symmetry) \label{qp}\enq
  but when the coupling strength are unequal a remarkable change occurs, as shown in the
  following figures (Fig. 1,2).

\begin{figure}
\vspace{6cm} \includegraphics{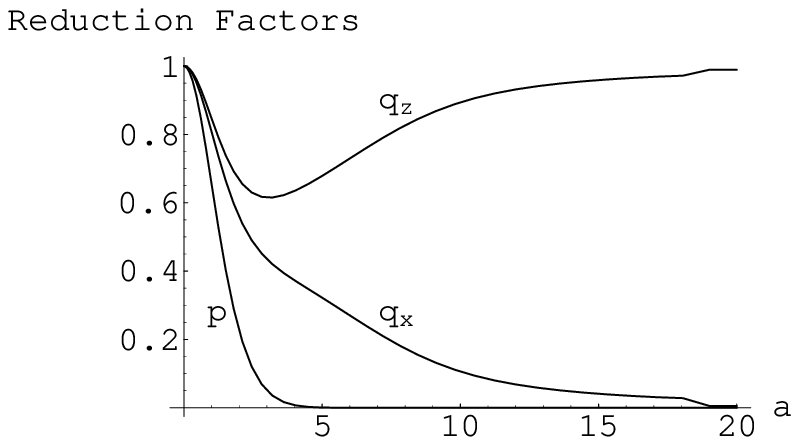} \caption {Reduction factors {\it
vs}. dominant coupling strength $a$. The diagonal $q_z$ and the two
off-diagonal reduction factors $q_x$ and $p$ are shown. Unlike the
cubic-symmetry case,
 when $q_x=q_z$, in the dihedral case shown here the two
 $q$-factors are  dissimilar, while $p$ remains qualitatively similar
 to that in cubic symmetry.}The ratio of coupling strengths $\frac{b}{a}$ is
 maintained a constant, $0.8$
\label {ellipticfig1}
\end{figure}

\begin{figure}
\vspace{6cm} \includegraphics{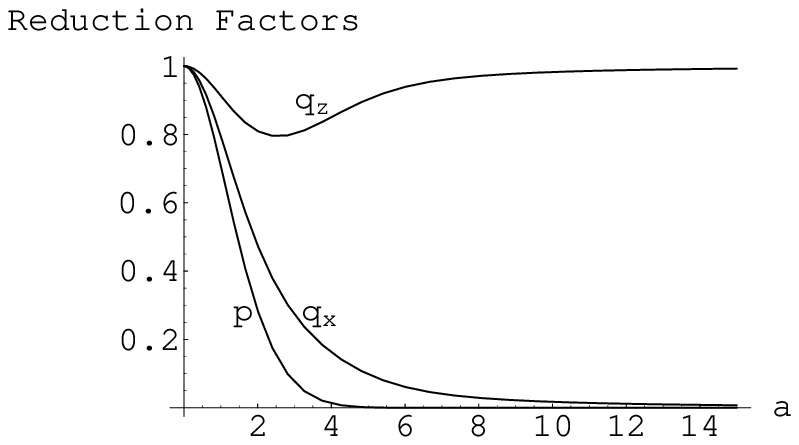} \caption {Reduction factors {\it
vs}. dominant coupling strength $a$.
 Same as figure 1, but with the ratio of coupling strengths $\frac{b}{a}$ decreased to $0.667$. The dissimilarity between
 $q_z$ and $q_x$ increases.}
\label {ellipticfig2}
\end{figure}

  These exhibit the three reduction factors as function of
   increasing\\ strength
  of the dominant coupling  and for two different values of the ratio $b/a$. It is apparent
  that while one $q$-factor decreases monotonically to zero (!), the other drops only slightly
  below unity and does so for only a limited range. (In cubic symmetry, also for linear coupling,
  the single $q$-factor decreases from unity to one-half.) The $p$-factor shows a  regular
   behavior.

  The following further results are of interest, and are capable of
  straightforward interpretation:

  ~(A)~ When (say) $b$ becomes very small, there is no "reduction" in $q_z$, which is $
  \approx 1$ except in  a small range of the strengths. In the limit of $b=0$ one has the
  simple polaron case, with no off-diagonal interaction.

   ~(B)~When (instead of the cases shown in the figure) $b>a$, the roles of the two reduction
    factors $q_x$ and $q_z$ are reversed.

    This can be justified as follows: Interchanging $a$ and $b$ in the
    Hamiltonian of \er{Hen} has the effect of interchanging the operators $\sigma_x$ and $\sigma_z$. This
    interconverts, per definition in \er{qx} and in \er{qz}, $q_x$ and $q_z$. Formally, the
    interchange can be performed  by applying
    the unitary transformation matrix $\frac{1}{\sqrt 2}(\sigma_x + \sigma_z)$ on the
     Hamiltonian.

   ~(C)~ The second relation in \er {qp}, which is expected to hold for linear coupling
   to a single-mode coordinate in O symmetry, does not in general hold for neither $q$
   separately, but holds accurately for their mean. Thus we find  under linear $b_1,b_2$
    coupling in dihedral symmetries that our data satisfy the new relations
    \beq p=q_x+q_z -1\label{pq}\enq

  ~(D)~Because the vibrational quanta
are absent from the (proposed, quasi-classical) vibronic wave-functions, they do not play
a role in the reduction factors, as long as the
 coupling constants are defined in the non-dimensional form, in the way done here.
 \subsubsection {Expressions for the reduction factors in dihedral symmetry}
 The following expression relates  the reduction factors to the $W$-coefficients
 \beq  W\left( \begin{array}{cc}
  a~b~c \\
  d~e~f
  \end{array} \right)\label{W}\enq defined and listed in \cite{Griffith} for point groups.
  (The definition of $W$ allows various rearrangements of the symbols, not detailed now.)

   For electronic states belonging to a doublet $E$
 \beq K(a)=(-1)^{(a)}\lambda(E) \sum_b (-1)^{(b)} W\left( \begin{array}{cc}
  a~E~E \\
  b~E~E
  \end{array} \right)<\chi_b^2>\label{Ka}\enq In the octahedral or dihedral groups $(-1)^{(a)}$
  is $-1$ for $a=A_2$ and $1$ otherwise. $\lambda(E)=2$ is the dimension of the $E$
  representation.  $<\chi_b^2>$ is the weight of the nuclear component having the $b$-representation
   in the vibronic state.
    The above formula is to be compared to the expression in
    Eq. (4.7.5) shown in \cite {BersukerP}, whose derivation is based on the Wigner-Eckart
  theorem, and to the corresponding expression for triplets in \cite {EnglmanCT}.

  Using \er{Ka} and Table D3.3 in \cite {Griffith} one arrives at the following expressions
  for the reduction factor in dihedral symmetry.
  \beq K(A_1)=1= <\chi_{A_1}^2> + <\chi_{A_2}^2>+<\chi_{B_1}^2>+<\chi_{B_2}^2>\label{KA1}\enq
\beq K(A_2)=p= <\chi_{A_1}^2> + <\chi_{A_2}^2>-<\chi_{B_1}^2>-<\chi_{B_2}^2>\label{KA2}\enq
\beq K(B_2)=q_z= <\chi_{A_1}^2> - <\chi_{A_2}^2>-<\chi_{B_1}^2>+<\chi_{B_2}^2>\label{KB2}\enq
\beq K(B_1)=q_x= <\chi_{A_1}^2> - <\chi_{A_2}^2>+<\chi_{B_1}^2>-<\chi_{B_2}^2>\label{KB1}\enq
\beq K(E)=0\label{KE}\enq
One notices immediately that $K(B_2)=q_z\ne q_x =K(B_1)$, as shown by the computed results in
Figures 1 and 2. Similarly, upon adding up the first two lines and the last two lines
 separately and subtracting the sums from each other, one obtains
 \beq  p -(q_x+q_z-1)= 4<\chi_{A_2}^2>\label{pq2}\enq
 This is similar to Eq 4.7.14 in \cite {BersukerP} obtained in $O$ symmetry. The right hand member
 is non-negative. However, for what are termed in \cite {BersukerP} "ideal cases", the right
  hand side
 is zero and one recaptures \er{pq}, as also found in our computation. Non-ideal cases are
 systems with coupling to more than one pair of modes \cite{HalperinE}, and others \cite
 {Napoleon,Fletcher}.

\subsection {Diagonalization within the {\it vibronic} doublet}\label{Diagonalization}
Two cases are of interest here. First, when the vibrational energies of the modes in \er{Hen}
are finite (this excludes acoustic modes in a solid), and the external fields are weaker than
the vibrational energies $|{\vec\Omega}|<< \omega_n $ (all $n$). Then the admixture by the external
 fields of higher vibronic states can be neglected and one can work within the ground state
 vibronic doublet. This is carried out here.

 Secondly, when the external field components are periodic with a  period $\frac{2\pi}
 {\omega_f}$ that is long, in the sense of $\omega_f <<|{\vec\Omega}|$, so that during a period
 the system will stay in the lowest vibronic state (the adiabatic theorem)\cite{CarolloFFV,
 WhitneyG}. Then, again, one can work
 solely within a ground vibronic state . This case will be treated in a future work.

In terms of the reduction  factors, the external field Hamiltonian in \er {Hef} changes, as follows:
\beq H_{el-f} =  \frac{\hbar}{2}{\vec \Omega }\cdot\vec{\sigma}\to \frac{\hbar}{2}{\vec
\Omega_v }\cdot\vec{\sigma_v}=H_{v-f}
\label{Hef2}\enq
where
\beq {\vec \Omega_v }=(q_x\Omega_x,p~\Omega_y,q_z\Omega_z)\label{Omegav}\enq
and $\vec{\sigma_v}$ are $2$x$2$ Pauli matrices defined in  the function space of the
{\it vibronic doublet}.

The constant interaction Hamiltonian $H_{v-f}$ has to be diagonalized between degenerate
eigenstates of $H_v$ in \er{Hv}. The diagonalization splits the states by an amount of \beq
 2\Delta = 2\hbar|{\vec
\Omega}_v|=2\hbar
R_{v} \label{splitting}\enq and selects the following two linear combinations of the $\theta$ and
 $\epsilon$ states:
 \ber |{\it l}> & = & -sin\frac{\theta_v}{2}e^{-i\frac{\phi_v}{2}}|\theta\}~ +~cos\frac{\theta_v}
 {2}e^{i\frac{\phi_v}{2}}
 |\epsilon\}\label{estate1}\\|{\it u}> & = & cos\frac{\theta_v}{2}e^{-i\frac{\phi_v}{2}}|\theta\}~
 +~sin\frac{\theta_v}{2}e^{i\frac{\phi_v}{2}}
 |\epsilon\}\label{estate2}\enr
 where we have defined two angles involving the three vibronic reduction factors and the
 three components of the field, through the expressions
 \beq \theta_v = arctan \frac{[(q_x ~\Omega_x)^2 + (p~\Omega_y)^2]^{\frac{1}{2}}}{q_z \Omega_z}
 \label{thetav}\enq
 and \beq \phi_v= arctan \frac{p~\Omega_y}{q_x~\Omega_x}\label{phiv}\enq
 (The reason for the notation is that ($R_v, \theta_v,\phi_v$) make up the spherical coordinate
 representation of the "reduced" field vector ${\vec \Omega_v}$ in \er {Omegav}.) $|{\it l}> $
  and $|{\it u}> $ are respectively the lower and upper split states of the doublet, but for
   $\Omega_z<0$ one has to take the branch between $\frac{\pi}{2}$ and $\pi$ in the inverse
   tangent $\theta_{v}$.

 \subsubsection{A counterintuitive effect of the off-diagonal coupling on the energy splitting}
 For any given strength of the  diagonal coupling, as the off-diagonal coupling increases,
 $q_z$ decreases. This is seen by comparing in figure 3 the three curves depicting
 $q_z$, (computed with $a>b$, so that the diagonal coupling is dominant) in which
 the curves decrease as the off-diagonal coupling strength increases. This
  behavior is unexpected for the following reason:
\begin{figure}
\vspace{6cm} \includegraphics{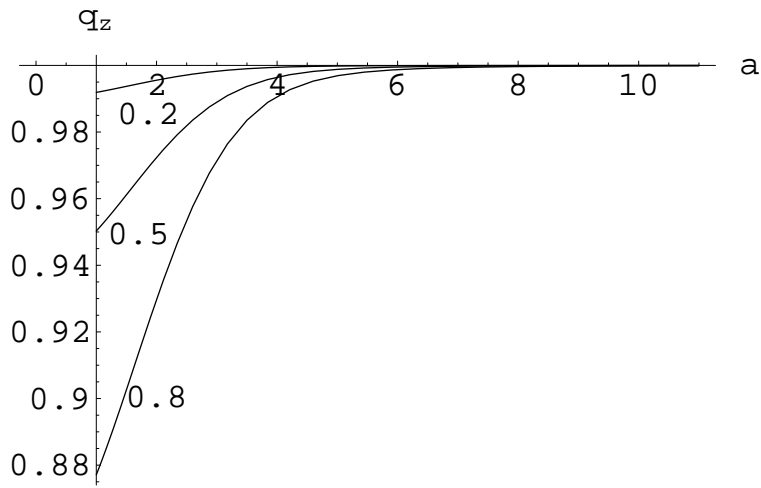} \caption {Dependence of the
diagonal reduction factor $q_z$ on the off-diagonal coupling
strength $b$. Reading the curves (here plotted against the
dimensionless diagonal coupling strength $a$) from above to below,
these were plotted for $b(<a) = 0.2,0.5,0.8$} \label
{ellipticfig3}
\end{figure}

 When $\Omega_x~=~\Omega_z~=~0$,  the  splitting of the vibronic  doublet components
 is, as we have just seen in \er {splitting}, $2|q_z\Omega_z|$. Here $\Omega_z$ is a
 constant, while the multiplier $q_z$ is a function of the coupling strength parameters.
  However, the often quoted phenomenon known as the "repulsion of neighboring energy
  levels by interaction between them" would seem to require that $q_z$ should
  {\it grow} as the off-diagonal coupling increases, so as to make the  splitting wider. This does not
  happen, and the reason is that the off-diagonal coupling $b_nq_{n}$ is not just a constant
  (a "magnitude"), but has a dynamical character. (We have purposely chosen for $b$ low values,
  to show that the result shown here is not a high-order effect in $b$. We also recall some
   related discussion in the literature that
 lay the claim that the "repulsion between levels" need to happen only when the set of
 states is complete.)
\section {Application to Some Low Dimensional Systems}
\subsection{Introductory remarks}
A number of organic hydrocarbon molecules or radical ions provide instances for a doublet interacting
with two independent modes. Investigations of their stereochemical properties (that is, their
possible configurations) in either the ground or excited states have been lucidly summarized
by Bersuker \cite {Bersuker}, giving also extensive references.

 When the molecules belong to a dihedral symmetry group, they can be subject to linear couplings by two types
  of non-totally symmetric  nuclear displacement modes. When these are of unequal strength,
   the adiabatic potential surface will have two minima along the dominant coordinate  and two saddle
points along the non-dominant one. [Couplings quadratic or of higher order in the mode
 coordinates (not considered in this paper and also not of paramount importance in many
 hydrocarbons) can turn the saddle points into minima \cite {Bacci}.] Further
 distortions from these simple configurations are also possible \cite
 {Bersuker, RoeselovaBJC}. Typical energy differences between alternative configurations are
  of the
 order of 2.5 kilo-kelvins, which are also the values in these systems for
  the Jahn-Teller (or stabilization) energies that are relevant to the considerations
   in this paper.

Transitions between minima take place typically across the saddle point (a "transition
 state");
this is discussed for the cyclobutane radical cation ($C_4H_8 ^{\cdot +}$) in \cite {JungwirthCB}.
(Figure 4.)

\begin{figure}
\vspace{6cm} \includegraphics{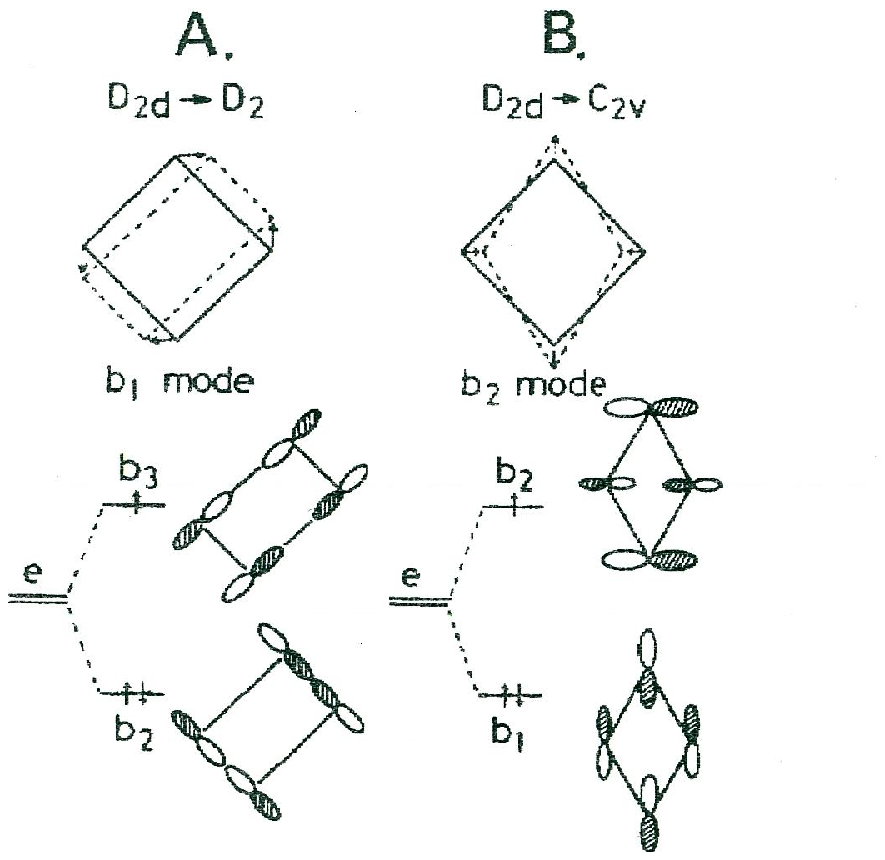} \caption{ Distortions and states of the
cyclobutane radical cation ($C_4H_8 ^{\cdot +}$). The upper part
shows two possible distorted configurations of the frame of the
four carbon atoms. Experiments favor the B configuration
($C_{2v}$). The puckered form of this configuration(two carbon
atoms above the paper and two below) is not shown;  neither are
the positions of the eight protons (also outside the plane of the
paper). The lower part shows the splitting of the energy levels of
the highest lying partly occupied doubly degenerate $e$-state
after distortion, as well as the shapes of the
 orbitals. Three spinning electrons are placed in the levels. The three-electron, product
state is labelled in the text $\theta$, while the state in which $b_2$ is doubly occupied and
$b_1$ singly occupied is labelled $\epsilon$.
Note that the electron densities on two-two carbon atoms differ in B, but are the same in A.
 (With permission, after \cite{UshidaSITN}. Copyright (1983) American Chemical Society.)}
 \label{ellipfig4}
  \end{figure}
The activation energies for the transitions are estimated in the same source as perhaps 1 kilo-kelvin,
possibly dropping below one half kilo-kelvin. With dynamic processes included, as in the present paper, the "transition" is of course an
 ingredient of the ground state wave function, (also termed "tunnelling") and does not represent
  a real process. Exceptions are when the molecule is embedded in a matrix with lower symmetry
  than the nominal one. Then the molecule may be forced into one of the minima, from which it can
  make a real transition to another minima. A further instance of real transition is a
  thermally activated one. The two states obtained in the last section, [\er {estate1},
  \er {estate2}] can be the
  basis for calculating transition rates but, typically, in a thermally activated process
  one includes states higher than the ground doublet and their inclusion is outside our concern
   here.

   Molecular data (like potential energy surfaces) have been calculated for the cyclobutane radical cation
   in \cite {JungwirthCB}, for the cyclobutadiene ($C_4H_4$) radical cation in \cite{RoeselovaBJC},
    and for
 radical cations of several cycloalkanes in \cite {OhtaNKS}. We cite \cite {Bersuker} for other systems.
   Calculated results tend to be very sensitive to the level of the computational effort (and
   this sensitivity
   also includes the order of relative stability among different configurations). In particular,
   single determinant wave-functions appear to be unreliable.

   Electron spin resonance (ESR) experiments have the capability of throwing light on dynamic
   effects, which we have studied here. Experimental determination of the vibronic reduction
   factors ($q_x,p,q_z$) from the observed $g$-factors in the spectra are unfortunately unlikely,
    due to covalency effects and the small
    spin-orbit coupling in these cyclic compounds ( this is unlike the transition ion compounds \cite
    {EnglmanH, Ham2}). However, the behavior of the proton hyperfine lines in hydrocarbons
    can give a clue to
   dynamic processes and, especially, to their coalescence and narrowing as the temperature
    is raised \cite {OhtaNKS,UshidaSITN}.
\subsection{ESR in the cyclobutane radical cation}
    In particular, we wish to apply the present theory to the hyperfine lines of
    $C_4H_8 ^{\cdot +}$ in a solid matrix observed and discussed by the authors of  \cite {UshidaSITN}. Their measurement of the electron
     spin resonance absorption at and above $77^0 K$ gave fairly equidistant nine-line spectra,
     which are
     indicative of electron-nuclear interaction with eight {\it equivalent}
     protons. On the other hand, low temperature observations at $4.2^0 K$ showed
     hyperfine lines with {\it different} separations. The observed separations are consistent
     with non-equivalent coupling to the four pairs of protons. This
      non-equivalence can be understood if one has a puckered molecule in which the
     four carbon atoms have undergone further distortion from $D_{2d}$ to  $C_{2v}$. This
     entails a $b_2$
     distortion-mode leading to a rhombus-like structure, as illustrated in figure 4.
     In our terminology,  for the coupling coefficients this implies that $|a|>|b|$.

     Part of the non-equivalence among the proton pairs may well disappear at higher temperature,
     either by the flattening of the puckered molecule or by thermal averaging between
     alternative non-planar forms (assuming a softening of the barrier in the matrix at the higher
     temperatures). However, the remaining distinctness between two pairs of carbons (located
     respectively on the blunt and the sharp angle apexes of the rhombus) cannot be fitted to
     the high-temperature isotropic spectra. It has been suggested in \cite {UshidaSITN}, that
     the dynamic Jahn-Teller effect is capable of removing this non-equivalence.

     Since the quasi-classical theory enables one to quantitatively evaluate the extent of
      "democratization" of the carbon atoms, we shall now provide some numerical estimates for
       this.
\subsubsection{Electron densities}
     The effects of Jahn-Teller distortions and of the motions between them on the hyperfine
     structure
     have been clearly formulated in \cite {McLachlanS, Snyder} and amply reviewed in \cite
     {Salem}, so that we do not have to repeat the theory.
     In essence, what one sees is the proton nuclear spin interacting with the electronic spin
      density on the carbons. For either of the carbon pairs,  $C_1$ and $C_3$ on the
       sharp angled
      apexes and  $C_2$ and $C_4$ on the blunt angled apexes, the spin density will differ in the
      $\zeta_{\theta}$ and  $\zeta_{\epsilon}$ electronic states in \er{zeta}. The
       density-difference of the two states will
      express itself in the spectrum in an opposite manner.

      On the other hand, the nuclear function cofactor of $\zeta_{\theta}$ and  $\zeta_{\epsilon}$ in the vibronic
      state will regulate the relative weights of these electronic states in the vibronic
      state. We are now looking for the average weights of these electronic states. They are clearly given
      by the expectation values of the projection operators $P_{\theta}$ and $P_{\epsilon}$
      in the vibronic state. These electronic projection operators can be written as
      \beq P_{\theta /\epsilon}= \frac{1}{2} (I\pm\sigma_z)\label{projector}\enq

      Suppose now that the host matrix of the cation radical stabilizes the
     $\zeta_{\theta}$ electronic state in preference to the $\zeta_{\epsilon}$ state. This
     stabilization can come about by having the following external field parameters : $\Omega_x=
     \Omega_y=0$ and $\Omega_z=-|\Omega^0|<0$, leading to
     \beq H_{el-f}=-\frac{\hbar}{2}|\Omega^0|\sigma_z \label {stabilization}\enq
     (Cf. \er{Hef}.) In the absence of any coupling to the nuclear motion the pure electronic $\zeta_{\theta}$
     state is stabilized.
     This state leads  clearly to non-equivalent couplings, whereas equivalence is regained only
    for states in which  $\zeta_{\theta}$ and $\zeta_{\epsilon}$ appear with equal weight.

     Let us now turn on the coupling to the nuclear coordinates.
     By the results in section 4.2, the lower energy solution is \er {estate1}. Since, with our
     choice  of the external fields, the angle
     $\theta_{v}=\pi$, this solution is simply the vibronic state $|\theta\}$
       shown in \er{vibronics}. We can now evaluate the expectation values of the electron
        projector operators in this state by using the definitions of the reduction factor
       $q_z$ in \er{qz}. We get \beq \{\theta|P_{\theta}|\theta\}=\frac{1}{2}(1+q_z)~~and~~
\{\theta|P_{\epsilon}|\theta\}=\frac{1}{2}(1-q_z) \label{projectors}\enq

What happens in the dynamic case? Remarkably, even here one does not achieve the extent of
 uniformity in  the spin density, which is achieved in the high symmetry case (e.g., $O$) for which
  $q_z=0.5$,
 since, as seen in figures 1 and 2, $q_z$ decreases only slightly
below 1, so that the difference between the two projectors will still remain. (Of course, this comes about because of
the dominance of the (rhombic)~$b_2$ distortion over the vying $b_1$ distortional mode.)
The end-result is that  under realistic conditions that the (rhombus-inducing) $b_2$
coupling is significantly stronger than the (rectangle-making) $b_1$ mode, the DJTE
 cannot make the two pairs of carbons equivalent, or the spectrum equidistant. For this
  to happen (at some elevated temperature), either the external splitting field $\Omega^0$ must be lowered,
  or fast jumps between the two vibronic eigenstates $|\theta\}$ and $|\epsilon \}$ or to
  higher lying vibronic states
   will have to occur. ("Fast" means shorter than $10^{-7}$ seconds, the hyperfine coupling
   time scale.)
\subsubsection{Dynamic effects on hyperfine lines}
   A numerical estimate for the $q_z$-factor confirms this conclusion. We have first
   estimated the coupling strengths for two coupling modes in the cyclobutane radical cation
   from
   computed stabilization energies in the rectangular and rhombic configurations, shown in Figure  7
   of \cite {JungwirthCB}. The values obtained from the UMP2/6-31G*//UHF/6-31G* variational method were
    adopted, since this places the rhombic configuration {\it below} ~the rectangular one, as
    is observed. The computed stabilization energies relative to the square configurations are
    $6600~cm^{-1}$ (for the rectangular shape) and $7250~cm^{-1}$ (for the rhombic form).
     Taking the experimental wavenumbers in cyclobutane  observed by \cite{LordN} for the
     modes: namely, $926~cm^{-1}$ in the $C-C$ stretching ($b_1$) and  $1001~cm^{-1}$ in
     the $CCC$ angle bending
    ($b_2$) modes, we obtain for the dimensionless coupling strengths: \beq
    a_{b_2}=7.611,~~~b_{b_1}=7.551
\label{coupling}\enq We note that these values are close to each other.

From our expression for the vibronic wave function and a simple quadrature, we obtain for
 these values a reduction factor \beq q_z = 0.999998
\label{coupling1}\enq that is, near enough unity, thus excluding the attainment of
equidistant spectra  by a pure dynamic mechanism alone.
 Phrased alternatively, the relevant matrix elements of the relaxation matrix are
proportional to the overlaps between Gaussian wave functions localized around different
 minima. These are greatly reduced by the strong vibronic coupling \cite{Polinger}.

 We recall that in $O$ symmetry, when the two coupling strengths $|a|$ and $|b|$ are precisely
  identical, so that  $q_z=q_x=q$, for strong linear coupling these reduction factors
   approach $\frac{1}{2}$, which indeed halves the difference between $\theta$ and
   $\epsilon$ occupancies. In $D_{2d}$ symmetry, when
   the coupling strengths $a$ and $b$ are different and large {$>>1$}, the tunnelling between
   different stabilized wells is negligible.

     \section{Summary and Outlook}
    We have tackled (fully, though not exactly) the time-independent quantum mechanics of
    a pair of isolated electronic
    states, and this under rather general conditions: namely, the states are subject to interaction
     both with static external fields
    and with a dynamic surrounding, the latter in the linear approximation.
   Our non-perturbational treatment was made possible (a) by having  a closed solution
   (the quasi-classical vibronic wave-function) for the part expressing
   the coupling between the electron and its dynamical surrounding, and
   (b) by inverting the usual order of solution through taking step (a) first and including
    the external field later. Thereby, the dynamically coupled states maintain convenient
    symmetry-group properties in the (Hilbert) function space; the external forces are subsequently
    treated within this framework. Their strength is, however, modified ("renormalized") by
     "vibronic" reduction factors. The use of these factors in a non-perturbational way is
     yet another new feature of this approach.

        Branching  out from the present treatment carried through for two states, one can
    similarly tackle problems with interactions affecting  three arbitrary electronic
   states, as well as any chosen number of states. By extension of the approach worked
  out in this work and shown in \er{Psihat}, one finds in these cases also that the wave-function
   generator, the generalization of $\hat{\Psi}$, is of a closed
   form. This consists of a finite number of terms, each with a given symmetry. Precisely,
  the generators (matrices) correspond  to all representations of the symmetric
   product of the electronic multiplet in its reference group. Thus, recalling the
   results in this paper, for a doublet we have
   {\it three} matrices (those appearing in \er{Psihat2}). (These three matrices are recognized
   as  representatives in $O$ of the symmetric product of the E-representation,
    namely $A_1$ and  $E$.) In a similar manner, for an electronic triplet
   one has {\it six} matrices: namely,  the identity matrix, three angular momentum matrices
    and two more trace-less matrices. All these make up the generator wave-function. Likewise,
     for any larger number of states. Algebraic relations connect the functions belonging
    to each term, which relations have to be solved simultaneously.
     (The point of these remarks is to assert that the doublet is not a fluke-case, but rather
     a special, though by far the simplest, case of multiple electronic states in interaction
      with their surroundings.)

In conclusion, we restate that the major restrictions on the applicability of this work and of
   its possible extensions
   are the validity of regarding a finite number of states in isolation and the linear
    approximation.

\section {Acknowledgments}
We thank Isaac B. Bersuker, Viktor Polinger and Boris Tsukerblat for help and encouragement
and  Serge Shpyrko for sending us the numerical data cited in section 3.4

\end{document}